\documentstyle[12pt,a4,epsf]{article}
\normalsize
\def\simless{\mathbin{\lower 3pt\hbox{$\rlap{\raise 5pt\hbox{$\char'074$}}
\mathchar"7218$}}}
\def\simgreat{\mathbin{\lower 3pt\hbox{$\rlap{\raise 5pt \hbox{$\char'076$}}
\mathchar"7218$}}}
\catcode`@=11
\def\beqra{\begin{eqnarray}} \def\eeqra{\end{eqnarray}}
\def\beq{\begin{equation}}      \def\eeq{\end{equation}}

\def\fo{\hbox{{1}\kern-.25em\hbox{l}}}

\def\ch{\@startsection{section}{1}{\z@}{-3ex plus-1ex minus-.2ex}%
        {2ex plus.2ex}{\large\sc}}

\def\; \lapp \;{\raisebox{-.4ex}{\rlap{$\sim$}} \raisebox{.4ex}{$<$}}

\def\con{\ifmmode \hbox{\bf*} \else{\bf*}\fi}   % conjugation
\def\scon{\ifmmode \hbox{\footnotesize\rm\bf*} \else{\footnotesize\rm\bf*}\fi}

\def\0#1{\relax\ifmmode\mathaccent"7017{#1}%    % puts a little circle atop,
        \else\accent23#1\relax\fi}              % as a halo of a saint

\def\eslash{\not{\hbox{\kern-2pt $E$}}}

\catcode`@=12
\begin{document}
\hoffset=-0.4cm
\voffset=-1truecm
\normalsize

%%%%%%%%%%%%%%%%%%%%%%%%%%%%

\begin{titlepage}
\begin{flushright}
%DESY ????\\
hep-ph/9504265
\end{flushright}
\vspace{24pt}

\centerline{\Large {\bf Particle Currents on a CP Violating Higgs Background}}
\vskip 12pt
\centerline{\Large{\bf and the Spontaneous
Baryogenesis Mechanism }}
\vspace{24pt}
\begin{center}
{\large\bf D. Comelli$^{a,}$\footnote{Email:
comelli@evalvx.ific.uv.es. Work supported by Ministerio de
Educacion y Ciencia de Espa\~{n}a}
, M. Pietroni$^{b,}$\footnote{Email: pietroni@vxdesy.desy.de}
 and A. Riotto$^{c,d,}$\footnote{Email:riotto@tsmi19.sissa.it.
Permanent address after November 95: Theoretical Astrophysics Group,
NASA/Fermilab, Batavia, Il 60510}}
\end{center}
\vskip 0.2 cm
\centerline{\it $^{(a)}$Departamento de Fisica Teorica,
Universitad de Valencia,}
\centerline{\it E-46110 Burjassot, Valencia, Spain}
\vskip 0.2 cm
\centerline{\it $^{(b)}$Deutsches Elektronen-Synchrotron DESY,}
\centerline{\it Notkestr. 85, D-22603 Hamburg, Germany.}
\vskip 0.2 cm
\centerline{\it $^{(c)}$Istituto Nazionale di Fisica Nucleare,}
\centerline{\it Sezione di Padova, 35100 Padua, Italy.}
\vskip 0.2 cm
\centerline{\it $^{(d)}$International School for Advanced Studies, SISSA-ISAS}
\centerline{\it Strada Costiera 11, I-34014 Miramare, Trieste, Italy.}
\vskip 2. cm
\centerline{\large\bf Abstract}
\vskip 0.2 cm
\baselineskip=15pt
We compute the particle currents induced on a bubble wall background at
finite temperature in a model with CP violation in the Higgs sector. Using a
field theory approach we show that fermionic currents arise at one loop, so
that a suppression factor ${\cal O}(h_t \phi/\pi T)^2$ with respect to
previous computations is found. The contributions to the Higgs currents are
also derived and their relevancy for the spontaneous baryogenesis
mechanism is discussed.
\end{titlepage}

\baselineskip=18pt
\setcounter{page}{1}
\setcounter{footnote}{0}
%\normalsize
The possibility of generating the baryon asymmetry of the Universe during the
electroweak phase transition has received much attention in the last years
\cite{krs, reviews}.
In the usual scenarios the transition is required to be first order and to
proceed via nucleation of bubbles of the broken phase in the unbroken phase.
The necessary departure from thermal equilibrium then can take place inside
or in front of the walls of the expanding bubbles.

Two different limits have been investigated in the literature. In the
case of thin bubble walls, the asymmetric (in fermion numbers)
reflection of particles off the
bubble wall is the dominant effect. The induced fermion number flux is then
reprocessed into a baryon asymmetry by the anomalous, $(B+L)$-violating,
sphaleronic transitions in the unbroken phase
\cite{reflection}.

In this letter we will focus on the opposite limit of thick bubble walls.
Indeed, this is thought to be
the relevant one if the phase transition
is weakly of the  first order, as it seems  to be the case for the standard
model
\cite{smpert,lattice} and the minimal supersymmetric extension of it (MSSM)
\cite{MSSM}.
If the only source of CP violation is the phase in the Cabibbo-Kobayashi-
Maskawa matrix, it seems very hard to generate any baryon asymmetry in this
limit. On the other hand, if CP violation is present in the Higgs
sector (which requires at least two Higgs doublets),  the so called spontaneous
baryogenesis mechanism can be invoked \cite{CKN}.
Both in the case of explicit \cite {CKN} and spontaneous \cite{scpv}
CP violation, a space-time dependent relative phase  $\delta(x)$ between the
two Higgs fields is turned on inside the bubble wall.
In order to analyze the effect of this complex space-time
dependent background on particle densities, a rotation on the fields
can be performed to make the Yukawa couplings real. As a consequence, a
derivative coupling of the form
 \beq
{\cal L}_{int}\sim \partial_\mu \delta J^\mu,
\label{deriv}
\eeq
where $J^\mu$
is the current corresponding to the rotation\footnote{In order to
avoid anomalies, in  the original paper \cite{CKN} and in \cite{diff}
the rotation was taken to
be the hypercharge. In fact, this can only be done if the same higgs
doublet couples both to up and down type quarks. In the models in which two
different Higgs doublets are coupled to the up and down
type quarks (as is the case
for the MSSM) it is not possible to remove
the relative phase $\delta$ from all the Yukawa couplings by making a
hypercharge rotation \cite{abel}.},
is induced from the kinetic terms.

In the original paper \cite{CKN}, only the time derivative of the phase
$\delta$ was taken into account. In this approximation, $\dot{\delta}$ acts as
an effective chemical potential (usually called `charge potential'),
 and the particle densities are perturbed to nonzero values
$n_i \sim q_i\: \dot{\delta}\: T^2$, where $q_i$ is the charge of the
$i$-th particle under the given rotation.
In presence of such a charge potential, and of baryon number violation, the
thermodynamical
evolution of the system adiabatically leads to a non-vanishing baryon
asymmetry. This approximation has been improved in ref. \cite{diff}, where
also the spatial derivatives of $\delta$ in eq. (\ref{deriv}) were taken
into account, and the role of particle diffusion was discussed.

Nevertheless, we believe that using the interaction term in eq.
(\ref{deriv}) as a starting point to compute the perturbations to the thermal
averages presents some problems.

First of all, the proportionality between an individual  particle density and
its hypercharge is
a direct consequence of the hypercharge rotation made to make the Yukawa
couplings real. Making a different rotation, the proportionality would of
course drop. For instance, one could rotate the right handed top and leave the
left handed one untouched, absorbing the anomaly by a proper rotation of the
light fermion fields. In this case, the
perturbations to be put in the system of kinetic equations describing the
system would be different, so that in principle the final value for the baryon
asymmetry might come out to be dependent on the rotation that has been made.

Secondly, since the phase $\delta$ is communicated from the Higgs to the
fermion sector through the Yukawa interactions, any perturbation in the fermion
densities $n_i$ should vanish in the limit of zero Yukawa couplings $h_i$.
Also, they
should vanish in the limit of zero vacuum expectation value for the Higgs
fields $H_i(x)=v_i(x)\exp[i\theta_i(x)]$
because no spontaneous CP violation is
present in the Higgs sector in
this limit. Naively, one could then expect a suppression factor of order
$(h_i^2  v^2_i(x)/T^2)$, where $h_i$ is the relevant Yukawa coupling,
for the perturbations in the fermionic particle number
 with respect
to the original result. Since we are interested in regions of
the bubble wall where
sphalerons are still active, {\it i.e} for values of $v_i(x)/T$  typically
smaller than
one, then the above mentioned suppression factor might be crucial.

The ultimate reason why these suppressions do not appear in the original
treatment, is that considering eq. (\ref{deriv}) as the only effect of the
background is equivalent to perturbing around the Higgs field configuration
 $\delta(x)=0$, $v_i(x)\neq 0$, which is not a solution of the field
equations. In other words, it is equivalent to disentangling $\delta(x)$ from
$v_i(x)$, whereas from the field equations one can  see that
$\partial_{\mu}\delta(x)$ vanishes as $v_i(x)^2$ for vanishing $v_i(x)$.

The purpose of this letter is to compute the  averages $n_i$ on the bubble wall
background, both for
fermions and Higgses, making use of a field theoretical approach. In this way
we
are able to treat the background consistently and to recover the expected
suppression factors. We wish to stress that what we call here $n_i$
are just the  perturbations induced by the CP violating background.
They should be used as source terms for the departure from equilibrium
in the equations describing
dynamical processes, like gauge and Yukawa interactions, baryon number
violation and particle diffusion.

Our starting point is the finite temperature
 generating functional for the
1PI Green's functions
with insertion of an operator $\hat{O}(z)$
(in the following $\hat{O}(z)$ will represent a
particle current)
\beq
\Gamma\left[\Phi^c_i(x),\:\Delta(x)\right] = W\left[J_i(x),\:\Delta(x)\right] -
\sum_j \int d^4x J_j(x) \Phi^c_j(x),
\eeq
where $\Phi^c_i(x)$ are the  classical
 fields of the theory and $J_i(x)$ the
corresponding sources, while $\Delta(x)$ is the source for the operator
$\hat{O}(x)$.

The quantity we are interested in is the expectation value of the operator
$\hat{O}(z)$ on the background given by the fields $\Phi^c_i(x)$, which we
will specify later,
\beq
\langle \hat{O}(z) \rangle_{\Phi^c_i(x)} =\frac{1}{i} \left.
\frac{\delta \Gamma\left[\Phi^c_i,\:\Delta\right]}
{\delta \Delta(z)}\right|_{\Delta=0}
\equiv {\cal O}\left[\Phi^c_i(x)\right](z).
\eeq
We can expand the functional ${\cal O}\left[\Phi^c_i(x)\right](z)$ in a
power series of $\Phi_i^c$
\beq
{\cal O}\left[\Phi^c_i(x)\right](z)=\sum_{n=0}^{\infty}
\sum_{i_1,\ldots,i_n} \frac{1}{n!} \int d^4 x_1\cdots d^4 x_n
{\cal O}_{i_1,\ldots,i_n}^{(n)}(x_1,\cdots,x_n;\:z)\Phi_{i_1}^c(x_1)\cdots
\Phi_{i_n}^c(x_n),
\label{exp}
\eeq
where the coefficients of the expansion are the n-point 1PI Green's
functions with one insertion of the operator $\hat{O}(z)$ computed in the
{\it unbroken} phase
\beq
{\cal
O}_{i_1,\ldots,i_n}^{(n)}(x_1,\cdots,x_n;\:z)=
\frac{1}{i}\left.\frac{\delta^{n+1}
\Gamma\left[\Phi^c_i,\:\Delta\right]}{\delta \Phi^c_{i_1}(x_1)\cdots
\delta \Phi^c_{i_n}(x_n)\delta \Delta(z)}\right|_{\Phi^c_i=\Delta=0}.
\eeq
The relevant background $\bar{\Phi}_i^c(x)$ for us is given by the bubble
wall, which is a  solution
of the field equations of motions
\beq
\left.\frac{\delta\Gamma\left[\Phi_i^c,\:\Delta=0\right]}{\delta
\Phi_i^c(x)}\right|_{\Phi_i^c=\bar{\Phi}_i^c}=0
\label{fe}
\eeq
with appropriate boundary conditions.

In order to be more specific, we consider the extension of the standard
model with two Higgs doublets where the up and down type quarks couple
to different Higgs doublets. We consider here the case of spontaneous CP
violation like that considered in ref. \cite{scpv} for the MSSM.

The most general tree level effective potential is given by
\begin{eqnarray}
V &=& {m_{1}}^{2}|H_1|^2 + {m_{2}}^{2}|H_2|^2 - ({m_{3}}^{2} H_1 H_2 +
{\rm h.c.})
+\lambda_1 |H_1|^4 +\lambda_2 |H_2|^4  \nonumber\\
&+& \lambda_3 |H_1|^2 |H_2|^2 + \lambda_4|H_1 H_2|^2
+\left[\lambda_5 (H_1 H_2)^2  + \lambda_6 |H_1|^2 H_1 H_2 +\lambda_7 |H_2|^2
H_1 H_2 + {\rm h.c.}\right],
\label{pot}
\end{eqnarray}
where all the couplings are now assumed to be real. Note that the potential $V$
depends only on the phase $\delta=\theta_1 + \theta_2$, whereas the orthogonal
combination represents the gauge phase.

We assume that the parameters of the Lagrangian and the temperature are such
that, when the loop corrections are considered, the potential assumes a double
well shape and a  expanding bubble solution exists for the complete equation of
motion given by eq. (\ref{fe}).

Since the interesting dynamics for baryogenesis takes place in a region close
to or inside the bubble wall, we approximate it with an infinite plane
travelling at a constant speed $v_w$ along the $z$ axis.

Considering eq. (\ref{fe}) at the tree level and including the  one loop
corrections to the effective potential, one can see that a solution exists
for which all the background fields are vanishing except for the neutral
Higgses. Their asymptotic behaviour is such that $v_i(z) \rightarrow 0 $ for
$z\rightarrow -\infty$ and $v_i(z)\rightarrow v^+_i$ for $z\rightarrow
 +\infty$, where $v^+_i$ are the finite temperature values for the Higgs
fields in the broken phase. Moreover, due to the dependence on the
phase $\delta$ of the potential, the two Higgs phases $\theta_{1,2}$ are
space-time dependent, and their evolution is such that
\beq
v_1^2(z) \partial^\mu \theta_1(z) = v_2^2(z) \partial^\mu \theta_2(z).
\label{fasi}
\eeq

This solution is the starting point for our
expansion in eq. (\ref{exp}), from which we will now see
that a non zero contribution
to the neutral Higgs currents already exists at the tree level, whereas
the currents for fermions and charged Higgses appear as loop effects. The
fact that we are not considering the complete background solution
to eq. (\ref{fe}) will modify our results only at higher orders.

We first consider the left handed top current $J^{\mu}_{t_L} =  \bar{t_L}
\gamma^{\mu}t_L$. The first non-vanishing term of the expansion in  eq.
(\ref{exp}) appears quadratically in the Higgs fields background, {\it i.e.}
\beq
\langle J^{\mu}_{t_L}(z) \rangle = \frac{1}{2}\sum_{i,j} \int d^4x\: d^4y
\left.\frac{\delta^{3}
\Gamma\left[\bar{\Phi}^c_i,\:\Delta_\mu\right]}{\delta \bar{\Phi}^c_i(x)
\delta \bar{\Phi}^c_j(y) \delta \Delta_{\mu}(z)}\right|_{\bar{\Phi}^c_{i,j}=
\Delta_{\mu}=0} \bar{\Phi}^c_i(x) \bar{\Phi}^c_j(y),
\label{quad}
\eeq
where we take $\bar{\Phi}^c_i=\{H_1^0,\: {H_1^0}^*,\:H_2^0,\:{H_2^0}^*\}$
to be the bubble wall Higgs background. Since we are interested in computing
the averages in regions where the sphalerons are still active, {\it i.e.} for
$v_i/T< 1$ \cite{dinethomas}, higher order contributions to the expansion
 (\ref{exp}) can
be safely neglected.

The first non-vanishing contribution
to eq. (\ref{quad}) appears only at one
loop, and is given by
\beq
\langle J^{\mu}_{t_L}(z)\rangle^{(1)}=
i\frac{h_t^2}{2}\int d^4x\:
d^4y \:{\rm Im}\left(H_2^0(x){H_2^0(y)}^*\right)\:
{\cal G}^\mu(x,y,z),
\eeq
where $h_t$ is the top Yukawa coupling and ${\cal G}^\mu(x,y,z)$ is the
Green function corresponding to the diagram in Fig. 1.
\begin{figure}
\leavevmode
\epsfxsize=5.cm
\epsfysize=5.cm
\centerline{\epsfbox{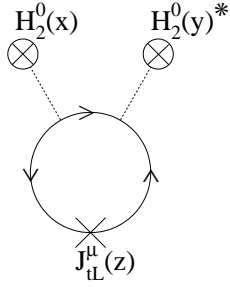}}
\caption{The 1-loop contribution to the left-handed top current.}
\end{figure}
 As we have already remarked, in computing
this diagram one must use the propagators in the unbroken phase. Note that
this contribution vanishes if the phase $\theta_2$ of the
Higgs field $H_2^0$ is a constant.
The scale of the external momenta $p$ are set by the space-time variation
of the phase, which, in the case of interest for us of thick
bubble walls ($L_w\simeq (10-100)/T$) \cite{thick}, allows us to neglect
contributions of order $p^2/T^2$ in the high temperature expansion.
Computing
${\cal G}^\mu(x,y,z)$ in the
$\overline{MS}$ renormalization scheme at the scale $\mu$, we obtain, in the
reference of frame of the thermal bath,
\beq
\langle J^{\mu}_{t_L}(z)\rangle^{(1)}\simeq -\frac{h_t^2}{2\:\pi^2}\:
v_2^2(z)\partial^{\mu}{\theta}_2(z)\:\left(\log\left(\frac{\mu^2}{A_f\:T^2}
\right) +\frac{7}{2}\right),
\label{jl}
\eeq
where $A_f=\pi^2 \exp(3/2 \:-2\gamma_E)\simeq 13.944$.

Eq. (\ref{jl}) shows the expected dependence on $h_t^2$ and $v_2^2(z)$
which, in comparison to the original result given in ref. \cite{CKN},
gives a suppression factor ${\cal O}(h_t v_2/\pi\:T)^2$.

A graph similar to that in Fig. 1 for the right handed top quark leads to a
contribution to $\langle J^{\mu}_{t_R}(z)\rangle$ given by $\langle
J^{\mu}_{t_R}(z)\rangle^{(1)} = - \langle J^{\mu}_{t_L}(z)\rangle^{(1)}$.
For the
other fermion species, one finds analogous results, in which $h_t$ is replaced
by the appropriate Yukawa coupling, and $v_2(z)$ ($v_1(z)$) appears for the up
(down)-type fermions.

A contribution to $\langle J^{\mu}_{t_L}(z)\rangle$ proportional to
${\rm Im} (H_1^0 H_2^0) = v_1 v_2 \sin\delta$, and to $h_t^2$,
appears at two loops, given by
the graph in Fig. 2. Since the computation has to be performed in
the unbroken phase, we must use resummed propagators for the Higgs fields in
order to deal with the IR divergences \cite{IR}.

In the unbroken phase the Higgs spectrum contains two complex neutral fields
and two charged fields.
The resummation can be achieved by  considering the propagators for the
eigenstates of the thermal mass matrix, which have masses given by
\beq
M^2_{1,2}(T) = \frac{m_1^2(T) + m_2^2(T) \mp
\sqrt{\left(m_1^2(T)-m_2^2(T)^2\right)^2
+ 4 m_3^4(T)}}{2},
\eeq
where the $m_i^2(T)$ are the thermally corrected mass parameters of the
potential (\ref{pot}), $m_1^2(T)\simeq m_1^2 + 3 g^2 T^2 /16$,
 $m_2^2(T)\simeq m_2^2 +  h_t^2 T^2 /4$, while $m_3^2(T)$ receives only
logarithmic corrections in $T$, which were computed in ref. \cite{scpv}.

Assuming $p^2\ll M^2_{1,2}(T) \ll (2 \pi T)^2$, where $p$ is again the
external momentum $p\simeq 1/L_w$, it is straightforward to see that the
non-zero Matsubara modes of the bosonic integral are strongly suppressed.
Keeping only the zero mode, we obtain
\beq
\langle J_{t_L}^{\mu}(z)\rangle^{(2)}\simeq -\frac{\lambda_5 \: h_t^2 }{384
\pi^3}
\frac{T\: m_3^2(T)}{(M_1(T)+M_2(T))^3}\left(\log\left(\frac{\mu^2}{A_f\:T^2}
\right) +\frac{7}{2}\right)
\partial^\mu \left[v_1(z)v_2(z) \sin \delta(z)\right].
\label{fatticunt}
\eeq
\begin{figure}
\leavevmode
\epsfxsize=5.cm
\epsfysize=5.cm
\centerline{\epsfbox{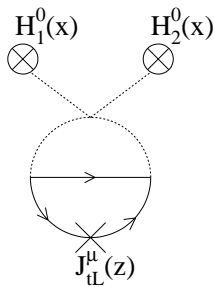}}
\caption{The 2-loop contribution to the left-handed top current. The scalar
internal lines can be either neutral or charged Higgs fields.}
\end{figure}
As for the one-loop result, the two loop contribution with neutral Higgses
in the internal lines for
$\langle J^{\mu}_{t_R}(z)\rangle$ is opposite to the one for
$\langle J^{\mu}_{t_L}(z)\rangle$.
This is no longer true when the graphs with charged Higgses in the internal
lines are taken
into account. In such a case, $\langle J^{\mu}_{t_R}(z)\rangle$ gets a
a contribution equal to eq. (\ref{fatticunt}), whereas
the result for $\langle J^{\mu}_{t_L}(z)\rangle$ is analogous but proportional
to $h_b^2$ instead of $h_t^2$.
The charged Higgs loops give also rise to a non-vanishing left handed bottom
density opposite to eq. (\ref{fatticunt}). As we will discuss in the
following, this fact may have
interesting implications for the spontaneous baryogenesis mechanism.

On the background of the CP violating bubble wall non-vanishing Higgs currents
\mbox{$J_{H_i}^{\mu} = i \left(H_i^{\dag} {\cal D}^{\mu} H_i -
{\cal D}^{\mu} H_i^{\dag} H_i\right)$}, where ${\cal D}^{\mu}$ is the
covariant derivative, are also present.
In the case of the neutral Higgses a contribution  already appears
at the tree level, see Fig. 3, and is given by the classical current
\beq
\langle J_{H_i^0}^{\mu}(z) \rangle^{(3)} = - 2 v_i^2(z) \partial^{\mu} \theta_i
(z).
\label{clas}
\eeq
Note that, again, the individual Higgs currents are zero if the phases are
constant. Moreover, as one can easily see from the equation of motion
(\ref{fasi}),
the total Higgs hypecharge current vanishes at the tree level.
\begin{figure}
\leavevmode
\epsfxsize=5.cm
\epsfysize=5.cm
\centerline{\epsfbox{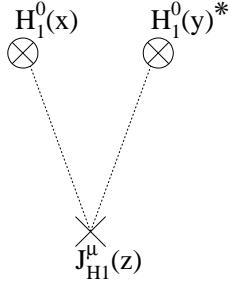}}
\caption{The tree level contribution to the neutral Higgs current.}
\end{figure}

At one loop, Fig. 4, we have a further contribution to the neutral Higgs
current and the first non-vanishing one to the charged Higgs current.
For the neutral Higgs currents we obtain
\beq
\langle J_{H_1^0}^{\mu}(z) \rangle^{(4)} =
\langle J_{H_2^0}^{\mu}(z) \rangle^{(4)} \simeq \frac{\lambda_5}{8 \pi}
\frac{T\: m_3^2(T)}{(M_1(T)+M_2(T))^3}
\:\partial^\mu \left[v_1(z)v_2(z) \sin \delta(z)\right],
\label{h1l}
\eeq
 Each charged Higgs gets a
contribution equal to that of the neutral Higgs belonging to the same doublet.
As expected, also the Higgs currents vanish in the limit of vanishing
$v_i(z)$.
\begin{figure}
\leavevmode
\epsfxsize=5.cm
\epsfysize=5.cm
\centerline{\epsfbox{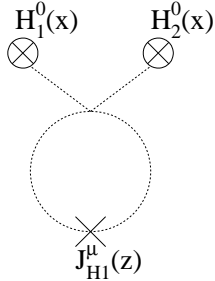}}
\caption{The 1-loop contribution to the neutral and charged Higgs currents.}
\end{figure}

In summary, we have made use of a field theoretical approach based on
a expansion in the background fields around the unbroken phase to
compute in a consistent way the
perturbations to the particle currents induced by a CP violating bubble wall
background.  We have shown that the various
contributions arise at the tree level, in the case of the Higgs currents, or
as loop effects. In this way it has been
possible to avoid the various ambiguities
inherent to the traditional approach based on the rotation of the
fields, recovering all the expected suppression factors and including  the
Higgs fields.

In order to get a feeling of the implications of these results for
the spontaneous baryogenesis mechanism, we can first consider an adiabatic
approximation similar
to that discussed in  ref. \cite{noi}. Inside the bubble walls the
$(B+L)$-violating sphaleron transitions are biased since a non-zero local
equilibrium value for $(B+L)$ is induced. $(B+L)_{EQ}$ is a linear
combination of the expectation values of all the currents conserved
by the interactions in equilibrium inside the bubble wall. Assuming
that gauge flavor diagonal, top Yukawa and Higgs-Higgs
interactions are in equilibrium \footnote{
In this approximation we also assume that
the hypercharge violating processes whose rates are
suppressed by powers of $v(T)/T$ are out of equilibrium.},
the conserved charges are $Q^{\prime}$,
$(B-L)^{\prime}$, $BP^{\prime}=B_3-1/2\:(B_1+B_2)$ and
$\tilde{Y}^{\prime}=Y_H+Y_{t_L}
+Y_{t_R}+Y_{b_L}+1/3\: Y^{\prime}_{lep}$ where the prime means that only
particles in equilibrium must be considered, and $Y^{\prime}_{lep}$ is the
hypercharge of the leptons in equilibrium. The values of these
conserved currents are made up by the background perturbations computed above.
Neglecting the contributions to the fermionic currents
not proportional to $h_t^2$, it is straightforward to
see that the contributions to $(B+L)_{EQ}$ from
the two Higgs
currents cancel each other both at the tree level and at one loop, and
 among the conserved charges
only $\tilde{Y}^{\prime}$
gets a non-vanishing  contribution at one loop.
The two loop diagrams where charged Higgses
are exchanged give a further contribution to $Q^{\prime}$ and
$\tilde{Y}^{\prime}$.

The local equilibrium value $(B+L)_{EQ}$ enters  a  rate equation of the form
\mbox{$\dot{B}\simeq -\Gamma_{SP} (B+L)_{EQ}/T^3$} where $\Gamma_{SP}$ is the
rate of the sphaleronic transition,
describing the generation  of baryon number inside the bubble walls.
As a consequence, the force driving baryon number violation is just
a one loop effect.

Regarding the final value for the baryon asymmetry in the
adiabatic approximation, the suppression terms
that we have found would give rise to a
suppression factor ${\cal O}(h_t^2 v_{co}^2/\pi^2 T^2)$
where $ v_{co}$ is
the value of the Higgs fields for which the sphaleron transitions
cease to be effective. Following ref. \cite{dinethomas} and taking $ v_{co}/T
\simeq \alpha_w/g$ we see that  typically  a  suppression
${\cal O}(10^{-4})$ arises. In this case, it might be hard to reconcile the
observed value for the baryon asymmetry with this mechanism for baryogenesis
in the adiabatic limit,
both in the case of spontaneous and of explicit CP violation in the Higgs
sector.

Nevertheless, it has been
recently shown that particle diffusion may play an
important role in the description of the spontaneous baryogenesis mechanism.
As a consequence of the perturbation induced by the bubble wall background,
and of the different diffusion coefficients of different particle species,
asymmetric densities are formed not only inside the bubble walls, as in the
adiabatic approximation, but also in front of it, where the
sphaleron transitions are not suppressed.  These asymmetries are then
transformed into a baryon asymmetry mainly in the region in front  of the
bubble wall, in a scenario similar to that occurring in
the case of thin bubble walls.

In order to improve the adiabatic approximation a system of kinetic equations
describing particle interactions and diffusion should then be solved, along
the way of, e.g., ref. \cite{diff}. In this case, the quantities
that we have computed would represent the sources of the departure from
thermal equilibrium.
This task goes beyond the scope of this letter, however
an important, and maybe helpful, difference with respect to the situation
considered in
ref. \cite{diff} may be pointed out.
As we have already discussed, in that
paper the perturbation to the i-th particle species was simply proportional
to $y^i \:\partial^{\mu}\delta \:T^2$, where $y^i$ is the
hypercharge of the particle. This proportionality was due to the fact that a
hypercharge rotation has been made to remove the Higgs phase from the Yukawa
couplings.  Such a dependence
does not appear in our results, neither for the fermions nor for the Higgses.
Instead, as long as the loops with
charged Higgses are neglected, we obtain that, e.g.,
$\langle J^{\mu}_{t_L}\rangle=-\langle J^{\mu}_{t_R}\rangle$. This behaviour
reflects the fact that in the Yukawa couplings left and right
handed quarks have opposite couplings to the imaginary part of
the Higgs field.  When charged Higgses are taken into account, the
contribution to $\langle J^{\mu}_{t_R}\rangle$ at two loops is opposite to
that to
$\langle J^{\mu}_{b_L}\rangle$ while $\langle J^{\mu}_{t_L}\rangle$ and
$\langle J^{\mu}_{b_R}\rangle$ get also opposite contributions, but only
proportional to
$h_b^2$. Concerning the Higgs currents, they are proportional to $\lambda_5$
instead of $h_t^2$ and moreover, from eq. (\ref{h1l}) we see that they
are the same for the two neutral Higgses, which have opposite hypercharges.

In ref. \cite{diff} it was noted that, since all the perturbations were
proportional to the hypercharge (and to $\dot{\delta}$),  the only processes
which could be biassed were the hypercharge violating ones, which have rates
suppressed by powers of $v_i(z)$. Due to the smallness of these rates on the
outer edge of the bubble wall, this gave rise to an important suppression on
the value of the fluxes of particles diffused in the unbroken phase.
On the other hand, our results do not exhibit such proportionality, then also
the (unsuppressed) hypercharge conserving processes are actually biassed.
As a consequence, inserting our perturbation in the kinetic equations one might
expect that in the unbroken phase fluxes of roughly the same magnitude than
those of ref. \cite{diff} would be found. A numerical treatment of the problem
using the results of this letter  and properly treating  baryon
number violation in the unbroken phase would then be highly desirable in
order to
understand whether diffusion can really help in overcoming the pessimistic
results of the adiabatic approximation.
\vskip 1.cm
It is a pleasure to thank W. Buchm\"uller, J.R. Espinosa, F. Feruglio, and
J. Ignatius for
useful discussions.


\newpage
\begin{thebibliography}{15}
\bibitem{krs} V.A. Kuzmin, V.A. Rubakov, and M.E. Shaposhnikov, Phys. Lett.
{\bf B155}, 36 (1985).
\bibitem{reviews} For a review see A. Cohen, D. Kaplan, and A. Nelson, Ann.
Rev. Nuc. Part. Sci. {\bf 43}, 106 (1993).
\bibitem{reflection} A. Cohen, D. Kaplan, and A. Nelson, Nucl. Phys.
{\bf B373}, 453 (1992), Phys. Lett. {\bf B294}, 57 (1992).
\bibitem{smpert} J.R. Espinosa, M. Quir\'os and F. Zwirner, Phys Lett.
{\bf B314}, 206  (1993);  W. Buchm\"uller, Z. Fodor, T. Helbig and D. Walliser,
Ann. Phys. {\bf 234}, 260 (1994); Z. Fodor and A. Hebecker, Nucl. Phys. {\bf
B432} 127 (1994);  W. Buchm\"uller, Z. Fodor and A. Hebecker,
hep-ph 9502321.
\bibitem{lattice} K. Kajantie, K. Rummukainen and M.E. Shaposhnikov,
Nucl. Phys.
{\bf B407} 356 (1993); K. Farakos,  K. Kajantie, K. Rummukainen and
 M.E. Shaposhnikov,  Phys. Lett. {\bf B336} 494 (1994);
B Bunk, E.M. Ilgenfritz, J. Kripfganz and A Schiller, Nucl. Phys. {\bf B403}
 453 (1993); F. Csikor, Z. Fodor, J. Hein, K. Jansen, A. Jaster, and I.
Montvay, Phys. Lett. {\bf B334} 405 (1994).
\bibitem{MSSM} J.R. Espinosa, M. Quir\'os and F.Zwirner,
Phys. Lett. {\bf B307}, 106  (1993); A. Brignole, J.R. Espinosa, M.
Quir\'os. and F. Zwirner, Phys. Lett. {\bf B324}, 181 (1994).
\bibitem{CKN} A. Cohen, D. Kaplan, and A. Nelson, Phys. Lett. {\bf
B263}, 86 (1991).
\bibitem{scpv} D. Comelli and M. Pietroni, Phys. Lett. {\bf B306}, 67
(1993); D. Comelli, M. Pietroni, and A. Riotto, Nucl. Phys. {\bf
B412}, 441 (1994); Phys. Rev. {\bf D50}, 7703 (1994);  Phys. Lett.
{\bf B343}, 207 (1995);
 J.R. Espinosa, J.M. Moreno, and M. Quir\'os, Phys.
Lett. {\bf B319}, 505 (1993).
\bibitem{diff} A. Cohen, D. Kaplan, and A. Nelson, Phys. Lett. {\bf B336}, 41
(1994).
\bibitem{abel}  S.A. Abel, W.N. Cottingham, and  I.B.
Whittingham, Nucl. Phys. {\bf B410}, 173 (1993).
\bibitem{dinethomas} M. Dine and S. Thomas,  Phys. Lett. {\bf
B328}, 73 (1994).
\bibitem{thick} M. Dine, R. Leigh, P. Huet, A. Linde, and D. Linde,
Phys. Rev. {\bf D46}, 550 (1992).
\bibitem{IR} P. Arnold and O. Espinosa, Phys. Rev. {\bf D47}, 3546 (1993).
\bibitem{noi}  D. Comelli, M. Pietroni, and A. Riotto, Padova preprint
 DFPD-94-TH-39, hep-ph/9406369.
\end{thebibliography}
\end{document}